\documentclass[preprint,superscriptaddress,amsmath,amssymb,prd,aps,showpacs,floatfix,nofootinbib]{revtex4-1}
\usepackage{graphicx}% Include figure filesa
\usepackage{dcolumn}% Aligntable columns on decimal point
\usepackage{bm}
\usepackage{mathrsfs}% bold math
\usepackage{hyperref}% add hypertext capabilities
\usepackage{amsmath}
\usepackage{amssymb}
\usepackage{bm}
\usepackage{color}
\usepackage{float}
\usepackage{dcolumn}
\usepackage{multirow}
\usepackage{changepage}
\usepackage{enumerate}
\hypersetup{pdftex,colorlinks=true,linkcolor=blue,citecolor=red,menucolor=black,urlcolor=blue,filecolor=blue}

\newcommand{\mev}{\textrm{ MeV}}

\begin{document}
%\title{$\boldsymbol{\Xi_{bb}}$ and $\boldsymbol{\Omega_{bbb}}$ molecular states}
\title{Inconsistency of the data on the $K_1(1270) \to \pi K^*_0(1430)$ decay width}
\date{\today}

\author{L.~Roca}
\email{luisroca@um.es}
\affiliation{Departamento de F\'{\i}sica, Universidad de Murcia, E-30071 Murcia, Spain}

\author{W.~H.~Liang}
\email{liangwh@gxnu.edu.cn}
\affiliation{Department of Physics, Guangxi Normal University, Guilin 541004, China}
\affiliation{Guangxi Key Laboratory of Nuclear Physics and Technology, Guangxi Normal University, Guilin 541004, China}
%\date{\today}% It is always \today, today,
%             %  but any date may be explicitly specified

\author{E.~Oset}
\email{oset@ific.uv.es}
\affiliation{Departamento de F\'{\i}sica Te\'orica and IFIC, Centro Mixto Universidad de
Valencia-CSIC Institutos de Investigaci\'on de Paterna, Aptdo.22085,
46071 Valencia, Spain}
\affiliation{Department of Physics, Guangxi Normal University, Guilin 541004, China}

\begin{abstract}
We show, using the same Lagrangian for the $K_1(1270) \to \pi K^*_0(1430)$ and $K^*_0(1430) \to K_1(1270) \pi$ decays,
that the present PDG data on the partial decay width of $K_1(1270) \to \pi K^*_0(1430)$ implies a width for $K^*_0(1430) \to K_1(1270) \pi$ decay
which  is about ten times larger than the total $K^*_0(1430)$ width.
A discussion on this inconsistency is done, stressing its relationship to the existence of two $K_1(1270)$ states obtained with the chiral unitary theory,
which are not considered in the experimental analyses of $K\pi\pi$ data.
\end{abstract}

%\pacs{11.30.Er, 12.39.-x, 13.25.Hw}% PACS, the Physics and Astronomy
                             % Classification Scheme.
%\keywords{Suggested keywords}%Use showkeys class option if keyword
                              %display desired
\maketitle

%\tableofcontents

%\section{Introduction}
%\label{sec:intro}

Data on $K\pi\pi$ produced in high energy diffractive $Kp$ and $Kd$ collisions
have been analyzed in the past and the $K_1(1270)$ and $K_1(1400)$ states were  identified more than forty years ago,
together with their decay channels \cite{CarnegieNPB,BowlerNPB}.
The $K^*\pi$ and $\rho K$ decay modes are the most prominent ones
but a surprisingly large experimental value for the branching fraction for the $K_1(1270) \to \pi K^*_0(1430)$ ($\pi \kappa$ in the past) appears.
A reanalysis of these data is done in Ref.~\cite{CarnegiePL} and the PDG \cite{PDG2020} quotes it as giving
\begin{equation}\label{eq:GammaPDG}
 \Gamma_{1}\equiv \Gamma [K_1(1270) \to \pi K^*_0(1430)]= 26\pm 6 ~{\rm MeV},
\end{equation}
for a $K_1$ with mass and width given by
\begin{equation}\label{eq:GammaPDG}
 M_{K_1}=1253 \pm 7 ~{\rm MeV}, ~~~~~~~ \Gamma (K_1(1270))= 90\pm 20~ {\rm MeV}.
\end{equation}
However, it is stated in the PDG that this partial decay width is ``not used for averages, fits, limits, etc."
On the other hand the only data not excluded for ``averages, fits, limits, etc." are those from Ref.~\cite{Daum},
with
\begin{equation}\label{eq:BRK1:2}
{\rm BR}[K_1(1270) \to \pi K^*_0(1430)]=(28\pm 4)\%,
\end{equation}
which is a big number as we shall see.

Furthermore, there is a much more recent experiment from Belle \cite{Guler},
which finds a significantly smaller branching ratio
\begin{equation}\label{eq:BRK1:3}
{\rm BR}[K_1(1270) \to \pi K^*_0(1430)]=(2.0\pm 0.6)\%,
\end{equation}
but, however, once again this datum is ``not used for averages, fits, limits, etc." by the PDG.
The summary data tables of the PDG give the number of Eq.~\eqref{eq:BRK1:2}.

In this short note we show that such a value is grossly inconsistent with the total width of the $K^*_0(1430)$
and the saturation of this width with the $K\eta$ and $K\pi$ decay channels,
with no trace of $K^*_0(1430) \to K_1(1270) \pi$ decay.

The quantum numbers of the $K_1(1270)$ are $I(J^P)=\frac{1}{2}(1^+)$ and for the scalar meson $K^*_0(1430)$ $\frac{1}{2}(0^+)$.
The transition from $K_1(1270) \to  K^*_0(1430) \pi$ with $\pi$ $1(0^-)$ requires a $p$-wave coupling to conserve angular momentum and parity.
This, together with the isospin coupling of a $\pi$ to two isospin $\frac{1}{2}$ structures
leads to the transition $t$-matrix from $K_1(1270) \to  K^*_0(1430) \pi$
\begin{equation}\label{eq:it}
-it={\cal{C}}\; \epsilon^\mu \; p_{\pi \mu} \, \vec{\tau} \cdot \vec{\phi},
\end{equation}
with $\epsilon^\mu$ the $K_1$ polarization vector, $\vec{\phi}$ the pion field in Cartesian basis
and $\vec{\tau}$ the Pauli matrix acting on spinors of isospin $\frac{1}{2}$ ($\vec{\tau} \cdot \vec{\phi}$
gives $\sqrt{2}$ for $K_1^{(+)} \to \pi^+ K^{*(0)}_0$ and $1$ for $K_1^{(+)} \to \pi^0 K^{*(+)}_0$).

The $K_1(1270) \to  K^*_0(1430) \pi$ decay width is given by
\begin{equation}\label{eq:GamK1:5}
\Gamma_{1}=\frac{1}{8\pi} \, \frac{1}{M_{K_1}^2} \, p_\pi \; \overline{\sum} \sum |t|^2,
\end{equation}
where $\overline{\sum} \sum |t|^2$ is the spin, isospin sum and average over the third components,
\begin{equation}\label{eq:sumt2}
\overline{\sum_{J{\rm -pol}}} \sum_{\rm isos} |t|^2=3\, {\cal{C}}^2 \, \frac{1}{3} \, \sum_{\rm pol} \, \epsilon_\mu\, p^\mu_\pi \, \epsilon_\nu \, p_\pi^\nu,
\end{equation}
and
\begin{equation}\label{eq:pol}
\sum_{\rm pol} \, \epsilon_\mu\, \epsilon_\nu =-g_{\mu\nu}+\frac{P_\mu P_\nu}{M_{K_1}^2},
\end{equation}
with $P^\mu$ the $K_1$ momentum.

For the $K_1$ at rest one finds
\begin{equation}\label{eq:GamK1:7}
\Gamma_{1}=\frac{1}{8\pi} \, \frac{1}{M_{K_1}^2} \, {\cal{C}}^2 \, p_\pi^3,
\end{equation}
with
\begin{equation}\label{eq:ppi}
p_\pi=\frac{\lambda^{1/2}(M_{K_1}^2, M_{K_0^*}^2, m_\pi^2)}{2M_{K_1}}\;\, \theta(M_{K_1}-M_{K_0^*}-m_\pi).
\end{equation}
Certainly Eq.~\eqref{eq:GamK1:7} only makes sense if the widths of the $K_1$ an $K_0^*$ are taken into account,
and the overlap of their mass distributions allows the $K_1(1270)$ to have mass components bigger than the mass of the $K_0^*$ plus a pion mass,
something not easy given the mass of the $K_0^*(1430)$,
but eased because of its large width.
From the PDG we have
\begin{equation}\label{eq:MKstar0}
M_{K_0^*(1430)}=1425\pm 50~{\rm MeV},~~~~~~
\Gamma_{K_0^*(1430)}=270\pm 80~{\rm MeV}.
\end{equation}
To take into account the mass distributions of the two resonances
we must convolve the width of Eq.~\eqref{eq:GamK1:7} with the spectral functions of the resonances:
\begin{equation}\label{eq:SR}
S_R(\widetilde{M})=\frac{-1}{\pi}\; {\rm Im} \frac{1}{\widetilde{M}^2-M_R^2+i M_R \Gamma_R}.
\end{equation}
Then we have
\begin{equation}\label{eq:Gam9}
\widetilde{\Gamma}_{1}=\frac{1}{N_1 N_2}\; \int d\widetilde{M}^2_{K_1} \int d\widetilde{M}^2_{K_0^*} \; S_{K_1}(\widetilde{M}_{K_1})
\cdot S_{K_0^*}(\widetilde{M}_{K_0^*}) \cdot \Gamma_{1}(\widetilde{M}_{K_1},\widetilde{M}_{K_0^*}),
\end{equation}
and $N_1, N_2$ are normalization factors used to account for some missing strength when integrating $d\widetilde{M}_{K_1}, d\widetilde{M}_{K_0^*}$
in some reasonable limits like $M_R\pm 2\Gamma_R$.
We have
\begin{equation}
  N_i=\int (-\frac{1}{\pi})\; S_{i}(\widetilde{M}_{i})\; d\widetilde{M}^2_{i}, ~~~(i=K_1, K_0^*)
\end{equation}
with the same limits for the integration as in Eq.~\eqref{eq:Gam9}.

On the other hand, we can use the same Eq.~\eqref{eq:it} to describe the $K^*_0(1430) \to K_1(1270) \pi$ decay,
which  
is  the time reversal reaction concerning the $K_i$ states.
In this case we evaluate $|t'|^2$ in the rest frame of the $K_0^*(1430)$ and we find
\begin{equation}\label{eq:sumt2New}
\overline{\sum} \sum |t'|^2=3\, {\cal C}^2 \; \left[ \frac{\left( M^2_{K_0^*}-m^2_\pi-M^2_{K_1}\right)^2}{4M^2_{K_1}}-m^2_\pi \right]
\end{equation}
and
\begin{equation}\label{eq:GamK0:5}
\Gamma_{K_0^*}=\frac{1}{8\pi} \, \frac{1}{M_{K_0^*}^2} \, p'_\pi \; \overline{\sum} \sum |t'|^2,
\end{equation}
with
\begin{equation}\label{eq:ppip}
p'_\pi=\frac{\lambda^{1/2}(M_{K_0^*}^2, M_{K_1}^2, m_\pi^2)}{2M_{K_0^*}}\;\, \theta(M_{K_0^*}-M_{K_1}-m_\pi).
\end{equation}
Once again we must use the convolution of Eq.~\eqref{eq:Gam9} to obtain $\widetilde{\Gamma}_{K_0^*}$
that takes into account the $K_1$ and $K_0^*$ mass distributions.

If we take into account the nominal masses and widths of the $K_1$ and $K_0^*$ of Eqs.~\eqref{eq:GammaPDG} and \eqref{eq:MKstar0}
and the nominal value of the $K_1$ width to $K_0^*(1430)$ of Eq.~\eqref{eq:BRK1:2} to obtain the value of the constant $\cal C$,
then we obtain
\begin{equation}\label{eq:GamK0toK1pi}
  \Gamma_0\equiv\Gamma_{K_0^*\to K_1 \pi^0}=2378~{\rm MeV}.
\end{equation}

This is a huge number, if not absurd, at odds with the total width of the $K_0^*$ of $270$~MeV.
The contrast is even bigger when we see in the PDG
that the width of the $K_0^*(1430)$ is practically exhausted with the $K\eta$ and $K\pi$ decays,
and there is no experiment having reported the $K^*_0(1430) \to K_1(1270) \pi$ decay.
We should note that even if we take the Belle results of Ref.~\cite{Guler} shown in Eq.~\eqref{eq:BRK1:3},
excluded ``for averages, fits, limits, etc." in the PDG,
the $K_0^* \to K_1 \pi$ decay width would be $170$~MeV, 
smaller than the total $K^*_0$ width,
but still incompatible with the fact that the $K\eta$ and $K\pi$ decays practically exhaust the $K_0^*$ decay width.

\begin{figure}[h]
\begin{center}
\includegraphics[width=0.6\textwidth]{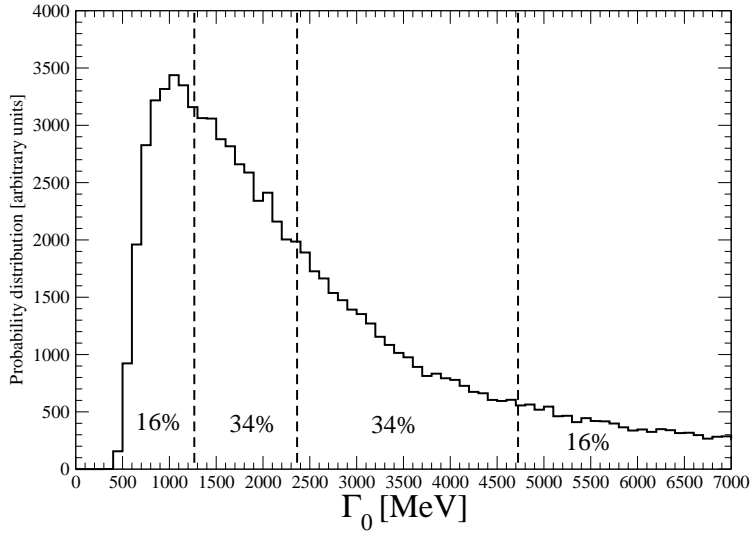}
\caption{\small{Probability distribution obtained for the $K^*_0(1430) \to K_1(1270) \pi$ decay width, $\Gamma_0$, by implementing a Monte Carlo sampling of the parameters within their errors. }}
\label{fig:hist}
\end{center}
\end{figure}

To further quantify the inconsistency of the PDG data on this partial decay width, $\Gamma_0$,
we carry out an error analysis taking into account all uncertainties of the different magnitudes.
This error estimation is also called for since the $K_1 \to K^*_0 \pi$ decay can proceed only from the overlaping of the spectral distributions in Eq.~\eqref{eq:Gam9} and then slight differences in the values of the parameters affecting the spectral distributions can lead to large differences in our prediction of the final $K_0^*\to K_1 \pi$ decay width. We perform a Monte Carlo sampling of the parameters in Eqs.~\eqref{eq:GammaPDG}, \eqref{eq:BRK1:2}  and \eqref{eq:MKstar0} within their errors and we find the probability distribution, $\rho(\Gamma_0)$, shown in Fig.~\ref{fig:hist}.

 We can see that the distribution of the $K_0^*\to K_1 \pi$ decay width is very asymmetrical, implying a highly nonlinear dependence on the parameters. Indeed, for many values of the random generated parameters there is none or very little phase space allowed for the  $K_1 \to K^*_0 \pi$  decay, which makes the predicted  $K_0^*\to K_1 \pi$ width to be  very large and then it moves much strength of the right tail of the probability distribution to high energies. Therefore we cannot provide a Gaussian error but rather we can summarize the probability distribution by means of other  statistical parameters like the {\it median} (the value with 50\% probability to the left and 50 \% to the right) (represented by the medium dashed line in  Fig.~\ref{fig:hist}), which is about $2386\mev$ and which essentially coincides with the value obtained in Eq.\eqref{eq:GamK0toK1pi} with the central values of the parameters. We can roughly assign a lower and upper  error to the previous value by considering the band of the width which encompasses 68\%  of the probability (see Fig.~\ref{fig:hist}) and then we have 
\begin{equation}\label{eq:GamK0toK1pierr}
  \Gamma_{K_0^*\to K_1 \pi^0}=2386_{-1284}^{+4720}\mev.
\end{equation}
The large value obtained for the upper error is again a consequence of the small region of the parameter space where the $K_1$ is able to decay into $K_0^*\pi$.
On the other hand the maximum of the distribution is at about $1000\mev$, but still incompatible with the experimental $K_0^*$ decay width. In addition, if we evaluate the {\it expected value}, or {\it mean}, of the $K_0^*\to K_1 \pi$ width, $\Gamma_0$, as $\overline \Gamma_0=\int{\Gamma_0\, \rho(\Gamma_0) d\Gamma_0}/ \int{\rho(\Gamma_0) d\Gamma_0}$, we get a much larger value,
$\overline \Gamma_0=4511\mev$, due to the large long tail at the right of the distribution as discussed above. 
If we use the Belle results of Eq.~\eqref{eq:BRK1:3} instead of 
the value in \eqref{eq:BRK1:2} we would get values of about $7\%$ of those quoted above, but again incompatible with the experimental total width of the   $K^*_0(1430)$ of $270\mev$ coming almost completely from 
$K\eta$ and $K\pi$ channels. The exact value of $\Gamma_0$ does not matter since we do not aim at providing an accurate value for it but to show the inconsistency of the 
$K_1(1270) \to \pi K^*_0(1430)$ quoted in the PDG.

On the other hand, we now recall that the PDG result of Eq.~\eqref{eq:BRK1:2} was obtained from the work of Ref.~\cite{Daum}.
The data of this work were reanalyzed in Ref.~\cite{LuisGeng}
to the light of the results of Ref.~\cite{LuisSingh}
in the study of the vector-pseudoscalar interaction with the chiral unitary approach,
where two $K_1(1270)$ states were obtained coupling mostly to $\rho K$ and $K^* \pi$ respectively (see also the review paper \cite{UlfRev}).
The data of Ref.~\cite{Daum} clearly showed the $\rho K$ and $K^* \pi$ distributions peaking at different energies,
but the analysis of Ref.~\cite{Daum}, redone in Ref.~\cite{LuisGeng},
obtained these structures from subtle interference of the amplitudes used in their analysis, which were model dependent.
It was shown in Ref.~\cite{LuisGeng} that the peaks observed experimentally were well reproduced by the two $K_1(1270)$ states picture.
The analysis of Ref.~\cite{Daum} also relied on the SU(3) mixture of the $K_1(1270)$ and the $K_1(1400)$ resonances that in Ref.~\cite{LuisGeng}
was discussed critically to the light of the existence of two $K_1(1270)$ states.

A revision and reanalysis of the data that led to the claim of the present PDG data for the
$K_1(1270) \to  K^*_0(1430) \pi$ partial decay width is necessary
and the new results of the Belle Collaboration \cite{Guler} seem to indicate that the official PDG results are grossly overcounted.
Yet, we believe that a final answer to this question will require an analysis
along the lines discussed in Ref.~\cite{LuisGeng} for the $\rho K, K^* \pi$ decay modes,
with the explicit consideration of the two $K_1(1270)$ states.

We think that it is important to pile up more experimental information supporting the existence of two $K_1(1270)$ states,
and take advantage to recall the suggestions made in the literature to attain this goal using the different reactions:
\begin{enumerate}[1)]
  \item $\tau \to \nu_\tau P^- K_1(1270)$, with $P^- \equiv \pi^-, K^-$ \cite{DaiRoca};
  \item $D^0 \to \pi^+ VP$, with $VP \equiv \rho K, K^* \pi$ \cite{Guanying};
  \item $D^+\to \nu e^+ VP$, with $VP \equiv \rho K, K^* \pi$ \cite{EnWang};
  \item $\tau$ decay to $\nu_\tau$ and two $K_1(1270)$ states \cite{Daimore};
   \item $\bar B\to J/\Psi  VP$, with $VP \equiv \rho \bar K, \bar K^* \pi$
 \cite{Dias:2021upl}.
\end{enumerate}

These and other reactions where $K\pi\pi$ is obtained in the final states,
separating the $\rho K$ and $K^* \pi$ modes,
will be most useful in the future to settle the issue of the two poles of the $K_1(1270)$
and at the same time resolve the problem of the flagrant inconsistency of the present PDG data on the $K_1(1270) \to  K^*_0(1430) \pi$ decay.

\section{ACKNOWLEDGEMENT}
This work is partly supported by the Spanish Ministerio de Economia y Competitividad
and European FEDER funds under Contracts No. FIS2017-84038-C2-1-P B
and by Generalitat Valenciana under contract PROMETEO/2020/023.
This project has received funding from the European Unions Horizon 2020 research and innovation programme
under grant agreement No. 824093 for the ``STRONG-2020" project.
This work is also partly supported by the National Natural Science Foundation of China under Grants No. 11975083 and No. 11947413.

%\clearpage

\end{document}